\documentclass[10pt]{article}
\usepackage[utf8]{inputenc}
\usepackage{graphicx}
\usepackage{amsmath,amssymb,amsfonts}
\usepackage{float}
\usepackage{color}
\usepackage{fullpage}
\usepackage{hyperref}
\usepackage{wrapfig}
\usepackage[affil-it]{authblk}
\usepackage{caption}
\usepackage{hyperref}
\usepackage{natbib}

\begin{document}

\title{Chain event graphs for assessing activity-level propositions in forensic science in relation to drug traces on banknotes}

\author[1]{Gail Robertson}
\author[2]{Amy L. Wilson}
\author[3]{Jim Q. Smith}
\affil[1]{Biomathematics \& Statistics Scotland, UK}
\affil[2]{University of Edinburgh, UK}
\affil[3]{University of Warwick, UK}

\maketitle

\renewcommand{\abstractname}{}

\begin{abstract}
\noindent Graphical models and likelihood ratios can be used by forensic scientists to compare support given by evidence to propositions put forward by competing parties during court proceedings. Such models can also be used to evaluate support for activity-level propositions, i.e. propositions that refer to the nature of activities associated with evidence and how this evidence came to be at a crime scene. Graphical methods can be used to show explicitly different scenarios that might explain the evidence in a case and to distinguish between evidence requiring evaluation by a jury and quantifiable evidence from the crime scene. Such visual representations can be helpful for forensic practitioners, the police and lawyers who may need to assess the value that different pieces of evidence make to their arguments in a case. In this paper we demonstrate for the first time how chain event graphs can be applied to a criminal case involving drug trafficking. We show how different types of evidence (i.e. expert judgement and data collected from a crime scene) can be combined using a chain event graph and show how the hierarchical model deriving from the graph can be used to evaluate the degree of support for different activity-level propositions in the case. We also develop a modification of the standard chain event graph to simplify their use in forensic applications. 
\end{abstract}

\section{Introduction}
Likelihood ratios are used in forensic science to evaluate scientific evidence and determine the extent to which this evidence supports competing propositions in criminal cases. The likelihood ratio of evidence $E$ under two competing propositions $H_1$ and $H_2$ is given by 
\begin{equation}
    LR= \frac{P(E \mid H_1)}{P(E \mid H_2)},
\end{equation}
i.e. the ratio of the probability of the evidence $E$ given $H_1$ divided by the probability of the evidence given $H_2$. The likelihood ratio (LR) can be used to quantify the level of support given by the evidence to the propositions (\cite{ref1}).

A crucial aspect of the likelihood ratio approach is the specification of the propositions used to evaluate the likelihood ratio. For example, in criminal cases competing propositions might be proposed by the prosecution and defence. The most basic type of propositions involve evaluating the extent to which evidence found at the crime scene has come from a particular source (e.g. whether DNA found at a crime scene could have come from a particular suspect). These are defined as source-level propositions (\cite{ref2}) and the simplicity of these statements makes calculating likelihood ratios relatively straight forward. While likelihood ratios associated with source-level propositions can be informative, for many types of evidence it is of more interest to consider propositions concerning whether or not a suspect performed a particular action (e.g. assaulted a victim, broke a window or sat on a car seat). Despite the utility of evaluating forensic evidence for activity-level propositions, in many cases these kinds of propositions are complex, requiring consideration of a suspect's activities and probabilities of transference and persistence of evidence (\cite{ref1}, \cite{ref3}, \cite{ref4}).  \par

To compare activity-level propositions new tools are needed which can account for complex chains of activities proposed by the prosecution and defense and so help frame the required likelihood ratios. Bayesian networks (BNs) have been used extensively to support probabilistic forensic analyses (see \cite{ref5}, \cite{ref6}, \cite{ref1}) and can be useful for evaluating activity-level propositions due to their ability to describe complex dependencies between different variables and compute likelihood ratios, and their ability to incorporate new evidence as it becomes available. However, for more complex activity-level propositions, BNs have some limitations. Firstly, a BN is not appropriate when the underlying event tree from which a BN is constructed is asymmetric (for example containing dead ends), as is commonly the case when representing evidence associated with activities graphically. Secondly, a BN does not display events temporally, which is a disadvantage when considering activity level evidence because it can be useful to display events as they occurred through time. As a result, when modelling scenarios such as the very asymmetric and time-ordered developments we can encounter with propositions in a criminal case it is natural to return to the more general setting of the probability or event tree to see whether these might provide a more suitable framework with which to assess activity-level evidence and so derive the corresponding likelihood ratios. However if we simply use a probability tree to represent a proposed scenario then we lose any conditional independence relationships that can be expressed within a BN, which are vital for inference. A chain event graph (CEG) embeds the developmental paths expressed in a probability tree and also captures conditionally independent relationships between nodes in these paths. This graphical structure has been successfully applied in diverse settings including as a support tool to help analysts detect crime (see e.g. \cite{ref7}). Although CEGs have been applied to illustrative toy examples in forensic analysis (see \cite{ref8} and \cite{ref9}), this paper is the first to explore the efficacy of this framework for translating the description of the propositions within a real criminal case and then deriving the likelihood ratios we need to analyse the strength of evidence in support of the prosecution case against the defence case. 
 \par

To construct a CEG we first draw a probability tree of the possible scenarios leading to the evidence to a level of granularity that can encompass all information and hypotheses relevant to  inferences about what actually happened in the case. Each root to leaf path therefore represents a particular version of events that explains what might have happened step by step. We next colour the non-leaf vertices (situations) and edges of the probability tree. Two vertices are assigned the same colour if the edges emanating from these  vertices are expected to share the same probability prior to any observations. Edges emanating from these different vertices are coloured the same if they have the same probability under this mapping. This produces a staged tree which embeds into the probability tree conditional independence statements, similar to those depicted in a BN. The CEG simplifies the staged tree by amalgamating two situations in which two coloured subtrees rooted at those vertices are isomorphic, and all leaf vertices are then drawn together into a single sink vertex. It is easy to check that this transformation preserves all the possible unfoldings of events represented in the probability tree by its root to leaf paths while often dramatically reducing the number of vertices and edges in the new graph. The CEG depicts directly both logical possibilities and context specific dependencies which are not depicted by the corresponding BN. Full details of the construction and properties of CEGs can be found in \cite{ref10} and \cite{ref9}. Bayes rule can be applied to a probability tree. Any direct observation of the process simply means that we have observed that a subset of the root to leaf paths has happened whilst its complement has not. We can therefore disregard all those edges in the complement (which will all have zero probability) and re-normalise path probabilities so that they all sum to one. Then the conditional probabilities on the edges of the residual subtree can be calculated from these atomic probabilities using the usual formulae. 

In this paper we modify the standard representation of a CEG so they can be better applied to forensic cases:
\begin{enumerate}
\item We depict the staged tree and CEG after conditioning on evidence which has been accepted by all parties in court so that we can focus only on storylines that are consistent with either the defence or prosecution case. To do this we draw the CEG posterior to what has been observed to have unfolded in the case, but the colours of the vertices and edges (and hence the edge probabilities) are kept as for the prior. This effectively sets to zero all other possibilities, greatly simplifying the depiction of the case and is what we do when our inference is led by the LR. This modification also allows us to represent evidence modelled as a continuous random variable.
\item We introduce two sink nodes in the CEG: one representing storylines that support the prosecution case and another that supports defence storylines (see for example \cite{ref11} for an analogous construction albeit for a very different setting).
\end{enumerate}

We argue that for cases involving activity-level propositions a staged tree and its associated CEG have many advantages. The storylines proposed by the prosecution and defence can be directly extracted from a CEG for use by a barrister to describe the unfolding of events in the case, which is compelling. The algorithms used in the CEG do not need the delicate construction of random variables to appropriately define any BN representation. Finally, the temporal coherence of represented storylines makes it easier to encourage jurors to implicitly introduce probabilities of events in a logical way, as these are described under the competing propositions. \par

We evaluate the use of CEGs for assessing activity-level propositions by considering an example based on a real-world drug-trafficking case (Compton and Ors v R. [2002] EWCA Crim 2835 (11 December 2002), focusing particularly on the case of Stephen Compton) in which potential evidence includes drug traces on banknotes. Cases involving evidence of drug traces on banknotes can be difficult to evaluate using LR methods as they typically involve sequences of activity-level propositions describing various possible ways in which the banknotes came to be contaminated. The LR in such cases is difficult to infer directly without considering the activities of the suspect or the origin of the notes, as various sources of drug transfer and their associated evidence need to be accounted for. In \cite{ref12} and \cite{ref13} methods for calculating LRs in such cases are explored for a specific pair of competing propositions but the authors noted that their proposed method would not be appropriate in cases which deviate from the specified propositions and may not extend to similar cases. CEGs allow LRs comparing prosecution and defence propositions to be calculated that combine various different sources of evidence (e.g. expert judgement and data collected from a crime scene) with descriptions of the activities that might have occurred. In this paper we demonstrate how this is done using as an example a series of storylines based on the Compton case. We describe how the CEG provides a new flexible framework for calculating LRs to evaluate degree of support for activity-level propositions using the case above as a real-world example . \\

\section{Example case}
\label{Ex1}

The judgement (Compton and Ors v R. [2002] EWCA Crim 2835 (11 December 2002)) provides information about an appeal in a criminal case against Stephen Compton (the suspect) which we have used to create an example case as described below which is representative of drug trafficking cases in general (e.g. contaminated banknotes and innocent activities which might explain the drug contamination). Stephen was one of three defendants in the case but for simplicity we have focused only on the case against him. The facts of the case that are agreed by the prosecution and defence are as below.
\\

The police seized $\pounds$107,000 of used banknotes stored in two safes at the suspect's address. The suspect was a drug user, although the type of drugs used was not specified. During the raid the police counted the banknotes on an uncovered table that may or may not have been contaminated with drugs. On seizure, the notes were transferred to tamper-evident bags which were also in contact with the uncovered table. The banknotes and tamper-evident bags were then sent to a forensic science provider for testing for contamination with heroin and other drugs. The tests revealed that a substantial proportion of the seized notes displayed traces of heroin. This proportion was greater than the proportion of notes contaminated with heroin in samples of banknotes taken at random from general circulation. An expert witness from Mass Spec Analytical (MSA) which carried out the forensic analysis gave evidence that 50\% of all notes seized in the case were contaminated compared with 2 to 5\% of notes in samples taken from general circulation. Of the ten tamper evident bags, MSA stated that only one was contaminated with heroin. In the original trial,  Stephen was convicted of possessing the proceeds of drug trafficking.

The evidence that we will model includes:
\begin{itemize}
    \item Measurements of drug contamination on the seized banknotes (heroin and other drugs),
    \item Measurements of drug contamination on ten tamper-evident bags,
    \item The suspect had $\pounds$107,000 stored in two safes at his address,
    \item The suspect did not bank $\pounds$107,000 cash.
\end{itemize}

Based on the judgement, we will consider two arguments that might be proposed by each of the prosecution and defence to explain the forensic evidence above and we will explore how to model the extent to which this evidence supports those arguments. Note that although the facts and evidence presented above are faithful to the court judgement, as this case is being used as an example of the use of chain event graphs for evidence evaluation we have adapted and embellished the arguments made in the judgement to better illustrate use of these methods. This is similar to the way in which CEGs might be used by lawyers to investigate possible arguments for explaining the forensic evidence and testimony in advance of a trial. Our analysis could be modified to include any alternative propositions by adding more root to leaf paths to depict further storylines.  We also wish to clarify that we use the example of the case against Stephen Compton only to demonstrate use of chain event graphs for forensic applications. We are not expressing an opinion on the merits of any arguments put forward by either side.

\subsection{Prosecution argument 1: Drug trafficking}

Banknotes which originated from general circulation were obtained by multiple drug-using individuals (from a bank, ATM or similar source) and the notes were contaminated with heroin while in their possession. The suspect obtained the contaminated notes from these individuals in exchange for heroin. After completing the drug deals, the suspect kept the contaminated banknotes at his address and did not bank them. The source of contamination was transfer of drugs to the banknotes while they were in the customers' possession.  

\subsection{Prosecution argument 2: Possessing the proceeds of drug trafficking}

Banknotes were obtained in return for the sale of drugs by an individual or individuals other than the suspect. The notes were contaminated with heroin while in the possession of these individuals or in the possession of their customers. The suspect obtained the contaminated notes from these individuals in exchange for jewellery. The suspect was part of the drug dealing operation and was fully aware that his customers were involved in dealing drugs and that this was the source of the notes.

\subsection{Defence argument 1: Suspect's drug use and/or police contaminated notes}

Banknotes with drug contamination consistent with that found on notes in general circulation were obtained by multiple non-drug-using individuals. The suspect obtained the banknotes from these individuals in exchange for jewellery via legitimate business deals. The suspect regularly handled the banknotes to swap new notes into the bundles for tax purposes and therefore contaminated the notes with drugs from his own habit. The table on which the notes were counted was also contaminated with drugs from his own habit. The source of drug contamination was transfer of drugs to the banknotes from the contaminated table as well as contamination due to the suspect's handling of the notes.

\subsection{Defence argument 2: Suspect unaware of notes' origins}

Banknotes were obtained in return for the sale of drugs by an individual or individuals other than the suspect. The notes were contaminated with heroin while in the possession of these individuals or in the possession of their customers. The suspect unknowingly obtained the banknotes from these individuals in exchange for jewellery via business deals. Heroin may have been transferred to the banknotes from the table during counting or from the suspect's own drug habit, but the fact that the notes were contaminated while in the possession of customers' involved in drug trafficking was the main source of heroin contamination on the notes.

\section{Formulating a chain event graph}

\subsection{The Evidence}
\label{section:evidence}

As described above, four pieces of evidence discussed in the judgement are relevant to our case:
\begin{itemize}
    \item E1 - The fact that the banknotes found at the suspect's address had not been banked or spent by the suspect, but were kept in safes at the suspect's address; 
    \item E2: The number of used banknotes found at the suspect's address (over $\pounds$100,000), an unusually large amount of money, suggests that a large profit was made from trading (illicitly or otherwise); 
    \item E3: The measurements of drug contamination on the banknotes found at the suspect's address ($\mathbf{Y}$). The contamination from drugs other than heroin found on each note may allow the source of the notes (e.g. multiple drug-using customers or the suspect himself) to be determined, i.e. if notes were contaminated by multiple customers we might expect to see contamination from other types of drug as well as heroin, but if the notes were contaminated by the suspect and if the suspect only used heroin we would expect to see background levels of other drugs on notes; 
    \item E4: The measurements of drug contamination on the ten tamper-evident bags placed on the table on which the notes were counted ($\mathbf{Z}$). 
\end{itemize}
Matrices $\mathbf{Y}$ and $\mathbf{Z}$ represent measurements of the quantity of different drugs (including heroin) on individual banknotes and tamper-evident bags retrieved from the suspect's address respectively. $\mathbf{Y}=\{y_{ij}\}$ is therefore a matrix of dimension $(n, m)$, where $m$ represents the number of drugs tested and $n$ gives the number of banknotes seized and $\mathbf{Z}=\{z_{ij}\}$ is a matrix of dimension $(10, m)$. 

Notes and bags were tested for drugs using tandem mass spectrometry, see \cite{ref14} for further details. The measurement associated with each banknote and bag for each drug is the $log_{10}$-transformed area under a peak, where the peaks are formed from counts of a particular product ion through time. For each drug two different product ions are measured and the ratio of these is used to identify the drug. As these two product ions are highly correlated we assume that just one is used as evidence for each drug. In the case of heroin this is the m/z 268 product ion. The drug measurements for the banknotes and bags are therefore continuous measurements. Later in the example we simplify this slightly and instead use a binary variable for the contamination on each tamper-evident bag for each drug, setting $z_{ij}=0$ when there is no contamination with that drug on that bag and $z_{ij}=1$ when there is contamination. With this approach, the forensic analyst will need to decide a level of contamination below which counts as zero (for example the limit of detection of the mass spectrometer). This would not be appropriate for the banknote measurements as it is known that most banknotes are contaminated with cocaine (\cite{ref12}, \cite{ref13}). 

\subsection{Topology of the Staged Tree and Chain Event Graph}

\begin{figure}
\includegraphics[width=\linewidth]{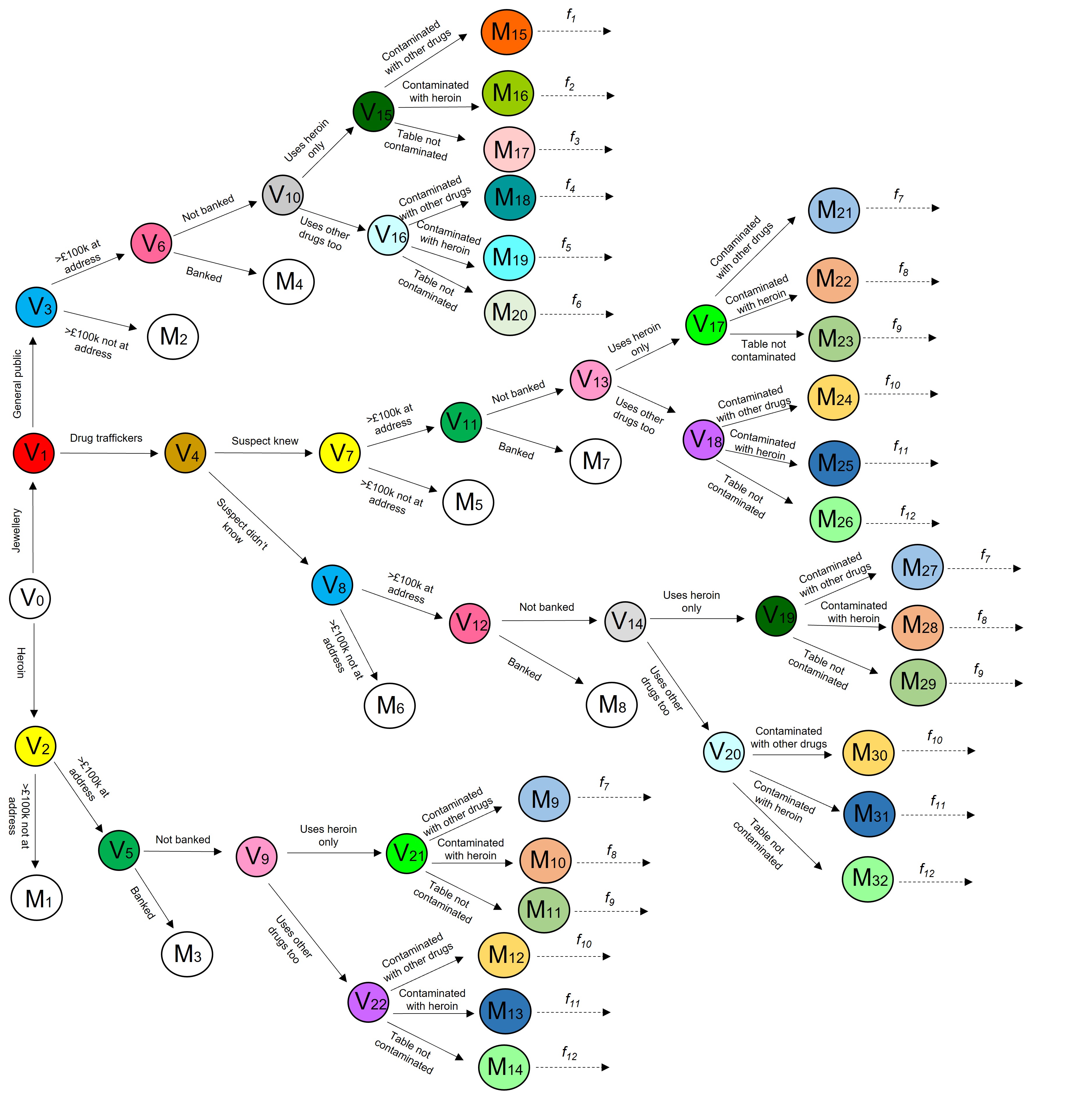}
\caption{Staged tree for example case where edges represent events in prosecution and defence storylines. Nodes with same colour represent situations where the emanating edges have the same probabilities. \label{ET1}} 
\end{figure}

The prosecution and defence each propose two possible hypotheses in our example case:
\begin{itemize}
    \item ${H}_{p_1}$: The suspect obtained the banknotes by dealing drugs 
    \item ${H}_{p_2}$: The suspect knowingly used a jewellery business to launder money obtained by the drug trafficking activities of others.
    \item ${H}_{d_1}$: The suspect had a legitimate jewellery business selling jewellery to the general public
    \item ${H}_{d_2}$: The suspect used a jewellery business to launder money obtained by the drug trafficking activities of others but the suspect was not aware of the source of the banknotes
\end{itemize}
The main events that occurred in the example case according to the prosecution and defence arguments are summarised in Table~\ref{table:Table1} and Table~\ref{table:Table2} respectively. We used these storylines of events to construct a staged tree as shown in Figure \ref{ET1}. The relationship between events in the example case and the staged tree is illustrated in the tables which show how the edges in the tree relate to different arguments. The colours in the tables also match those in the nodes. Where two nodes are the same colour, this means that the edges emanating from the nodes are associated with the same probabilities. For example in Figure~\ref{ET1} $V_6$ and $V_{12}$ are the same colour because we expect the probability that the suspect banked the notes to be the same under both ${H}_{d_1}$ and ${H}_{d_2}$ (as he did not know that the money was associated with crime). 

The final edge in the staged tree represents the forensic evidence $E_3$ and $E_4$, consisting of the continuous measurements of drugs on the banknotes and the tamper evident bags. The label $f_i$ on these edges for $i \in \{1, \ldots, 12\}$ represents the probability density function associated with this evidence, conditional on the preceding edges in the graph. 

The judgement does not specify which drug the suspect uses, but as drug use is given as a possible explanation for the heroin contamination of the table and hence banknotes we will assume that the suspect is a heroin user. We therefore include edges in which the suspect uses heroin only and in which he uses multiple different types of drugs (including heroin). We also include different scenarios of possible table contamination - heroin only, multiple drugs including heroin and no heroin contamination. In the example later in the paper we only model the heroin contamination on the banknotes and the bags hence we have not included edges here relating to contamination of the table with drugs other than heroin. To model other drugs too, extra edges would need to be added to the staged tree here to specify which drugs were being considered. 

We recognise that probabilities of transfer and persistence of drug traces may be relevant to some of the arguments in the staged tree. Some of these probabilities are implicitly encoded in the edges of the staged tree, for example the probability that a table is contaminated given that the householder is a drug user. For simplicity however we do not consider these in great detail (for example we do not consider timescales associated with contamination), but the staged tree could be expanded to include such considerations. 

\begin{figure}[H]
\includegraphics[width=\linewidth]{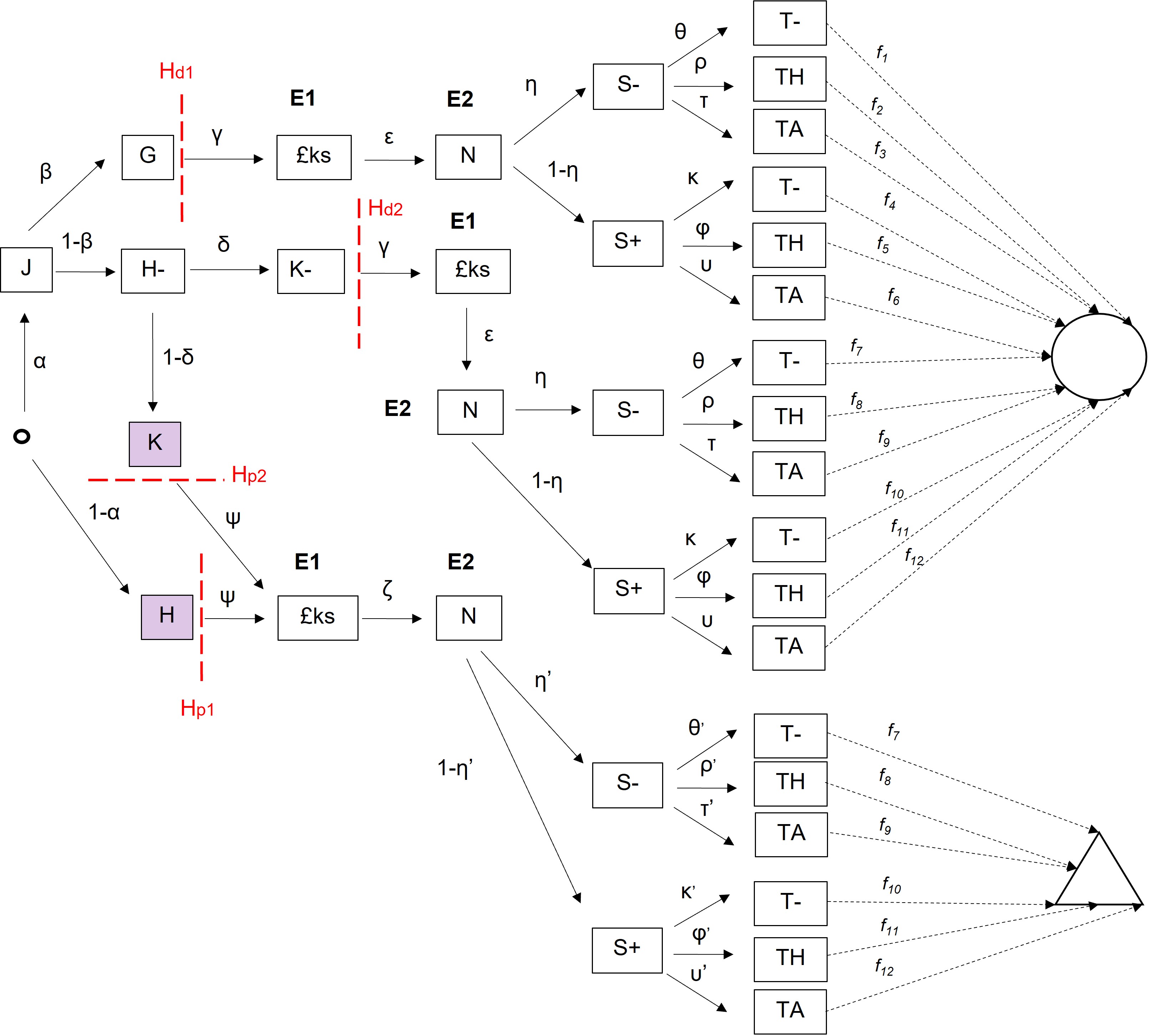}
\caption{Chain event graph for example case. Prosecution and defence hypotheses are shown by dashed red lines. Greek letters represent event transition probabilities. Storylines ending at the circle represent those proposed by the defence and storylines ending at the triangle represent storylines proposed by the prosecution. \label{ET2}}
\end{figure}

The chain event graph in Figure \ref{ET2} is formed from the staged tree in Figure \ref{ET1} by merging situations from which the two coloured subtrees rooted at those vertices are isomorphic (i.e. where all the edges following a given vertex have the same topology and the vertices are the same colour). For example, $V_5$ and $V_{11}$ in the staged tree are merged to form the lower part of the CEG from $N$. 

To make the CEG easier to read we have changed the vertex labels to better reflect the events that the edges leading into the vertices are representing. The following nodes included in the CEG represent the main events argued in the prosecution and defence storylines:
\begin{itemize}
\item $J$ source of banknotes is sale of jewellery by suspect 
\item $H$ source of banknotes is sale of heroin by suspect 
\item $H^{-}$ suspect is using jewellery business to launder money either knowingly or unknowingly
\item $G$ suspect is not using jewellery business to launder money and thus is selling jewellery to the general public
\item $\pounds ks$ suspect has $>$ \pounds 100k+ at address, $\boldsymbol{E2}$
\item $N$ notes were not banked, $\boldsymbol{E1}$
\item $K$ suspect knew the source of notes and was part of a drug trafficking operation 
\item $\overline{K}$ suspect did not know the source of the notes 
\item $S^{-}$ suspect used heroin only
\item $S^{+}$ suspect used heroin as well as other drugs
\item $T^{-}$ table on which notes were counted was not contaminated with heroin 
\item $TH$ table on which notes were counted was contaminated with heroin only 
\item $TA$ table on which notes were counted was contaminated with heroin and other drugs 
\item $\bigcirc$ the sink node for the defence storylines
\item $\triangle$ the sink  node for the prosecution storylines
\end{itemize}

Each prosecution storyline $s_{p_i}$ in the CEG is defined as a root-to-leaf path representing a sequence of events in the example case according to the prosecution. All prosecution storylines begin with edges representing one of the initial prosecution propositions $H_{p_j}$ for $j = 1, 2$. Defence storylines $s_{d_i}$ are defined in a similar way, beginning with the initial defence hypotheses $H_{d_j}$ for $j=1,2$. Cuts representing the end of the initial propositions are highlighted with dashed lines in the CEG (Figure \ref{ET2}). Event transition probabilities are represented as edges in the chain event graph. 

As for the staged tree, the final edge in the graph represents the forensic evidence $E_3$ and $E_4$. The probability associated with the edge is the probability density function of this evidence conditional on the preceding storyline evaluated at $\mathbf{Y}$ and $\mathbf{Z}$. For example, $f_7$ is shorthand for the probability density function $f(\mathbf{Y}, \mathbf{Z} \mid s_{d_7})$ which we have assumed is equal to $f(\mathbf{Y}, \mathbf{Z} \mid s_{p_1})$ and $f(\mathbf{Y}, \mathbf{Z} \mid s_{p_7})$ (hence all three are labelled $f_7$). 

We have made two modifications to the standard presentation of the staged tree and the CEG to better suit forensic applications. Firstly, both are presented posterior to what has been observed to unfold but the associated node colours are kept as for the prior. This is key to including evidence in the CEG. For example, in the staged tree in Figure \ref{ET1} the edge between nodes $V_6$ and $V_{10}$ represents the fact that the money was not banked. A priori, the money might have been banked (e.g. $V_6$ to $M_4$), but as we know that this did not happen and so we do not need to consider it, the edge is deleted from the staged tree and subsequent CEG. The probability associated with the edge from $V_6$ and $V_{10}$ is still our expectation that the money would not have been banked, conditional on the preceding events in the CEG, allowing us to compute LRs and other statistics. This modifcation is also key in allowing us to represent evidence based on continuous random variables in the CEG. The final edge in the staged tree and the CEG represents $E_3$ and $E_4$, i.e. the measurements of contamination on the notes and the tamper evidence bags. The probability associated with this edge is given by a continuous probability distribution -- the probability of these measurements given the storyline up to that point. Without the modification we would not be able to draw this evidence in the CEG as there are infinite possibilities. 

The second modification is to group the root to leaf paths in the CEG into two sink nodes rather than one - one for the prosecution storylines and one for the defence storylines. The reason for this is twofold. Firstly it makes the CEG more intuitive to those used to the adversarial legal system by clearly separating the two sets of arguments. Secondly, when computing the LR we need to keep defence and prosecution paths separated even if following the formal definition of the CEG they should be merged into one. This is because to compute the LR the prosecution storylines are included in the calculation for the numerator and the defence storylines are included in the calculation of the denominator (see section \ref{section:Bayesian} for more details). 

\begin{table}
\caption{\label{tab:Table 1} Table showing main events in the example case according to the prosecution (Prosecution case 1 and Prosecution case 2) and nodes in the staged tree associated with these events. Colours represent event stages with the same probabilities. Hypothesis 1 (${H}_{p_1}$: ${V}_0$ to ${V}_2$) and Hypothesis 2 (${H}_{p_2}$: ${V}_4$ to ${V}_7$) are highlighted in red} 
\centering
\includegraphics[width=\linewidth]{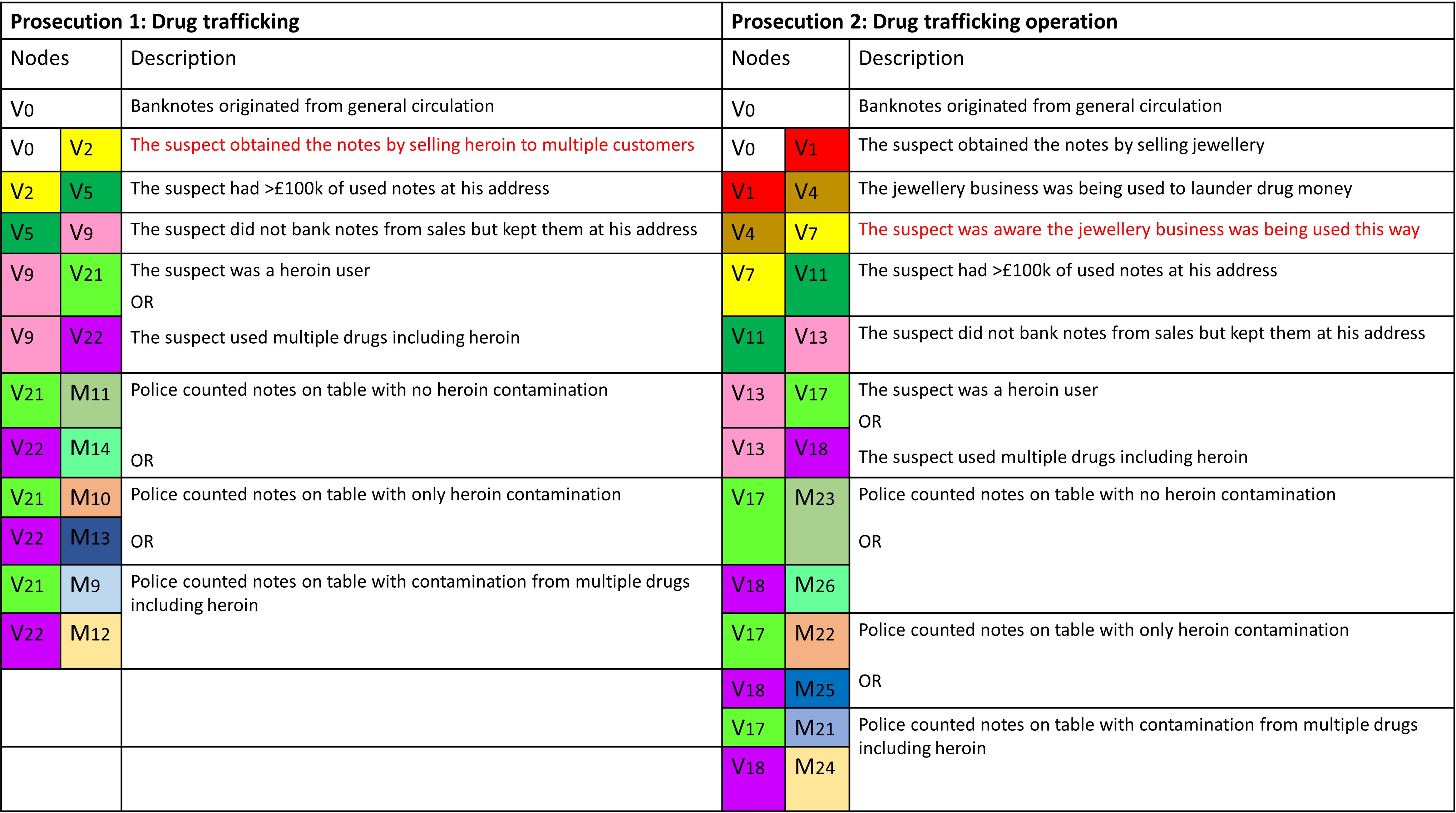}
\label{table:Table1}
\end{table}

\begin{table}
\caption{\label{tab:Table 2} Table showing main events in the example case according to the defence (Defence case 1 and Defence case 2) and nodes in the staged tree associated with these events. Colours represent event stages with the same probabilities. Hypothesis 1 (${H}_{d_1}$: ${V}_1$ to ${V}_3$) and Hypothesis 2 (${H}_{d_2}$: ${V}_4$ to ${V}_8$) that may have been proposed by the defence in this case are highlighted in red} 
\centering
\includegraphics[width=\linewidth]{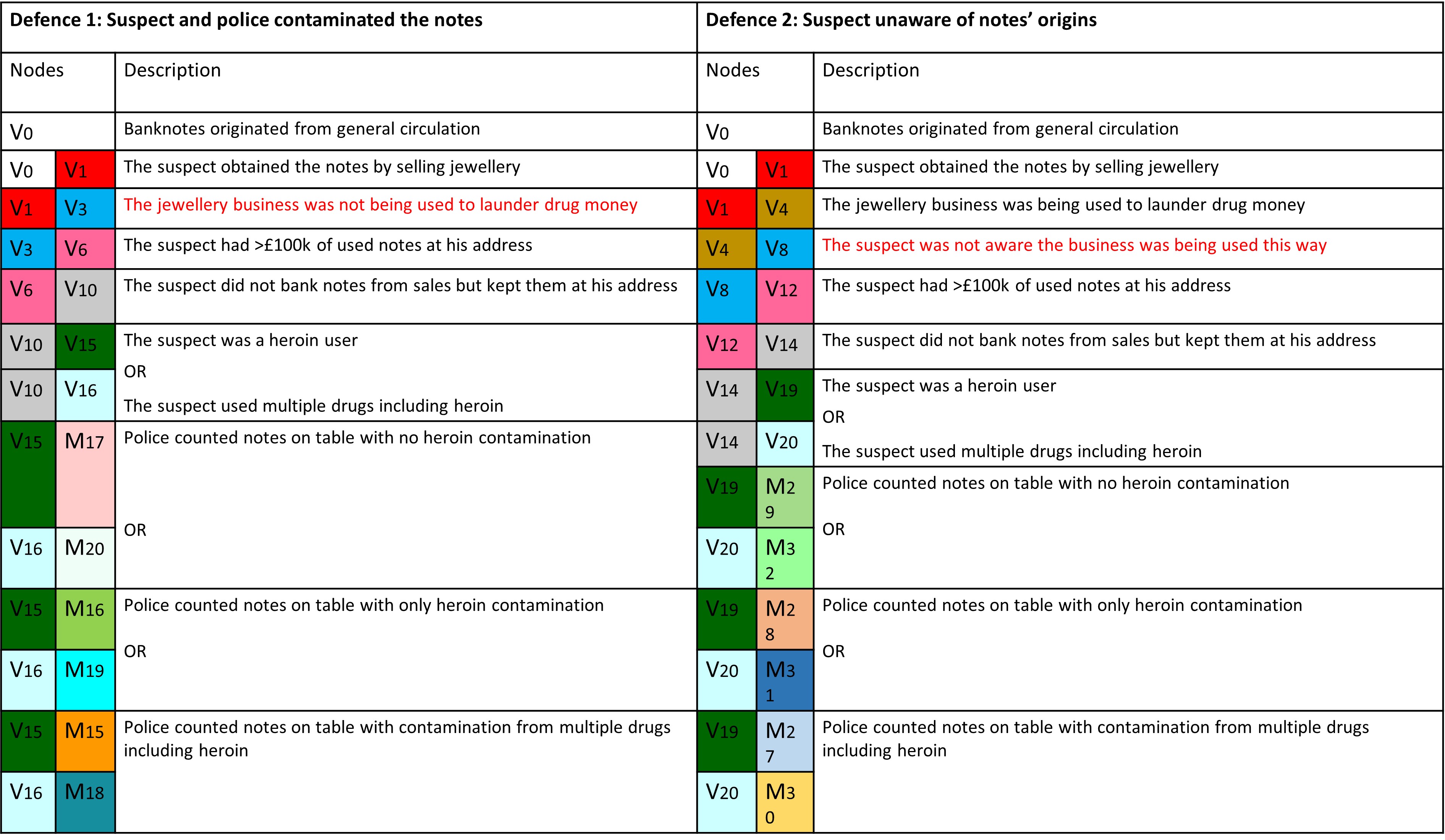}
\label{table:Table2}
\end{table}

\subsection{Determining the edge probabilities}
\label{section:key vectors}

The CEG is a mental map of arguments created by an individual. It is subjective and another individual might create a different CEG given the same information. The edges of the CEG refer to transition probabilities between events - these probabilities need to be set if the CEG is to be used to carry out probabilistic inference. The colouring of the staged tree also requires judgements to be made about whether the probabilities associated with different edges are the same or different. 

For the Compton case there are some datasets that might be used to inform the edge probabilities in the CEG:
\begin{itemize}
    \item Datasets of measurements of drugs on banknotes seized from convicted drug traffickers (e.g. see \cite{ref12}). One problem with such datasets however is that the measurements themselves may have been used as evidence to convict the individuals. A dataset in which the suspects plead guilty may be more representative.
    \item Datasets of measurements of drugs on banknotes collected from random samples of banknotes from general circulation (\cite{ref12}, \cite{ref15}, \cite{ref16}).
    \item Datasets consisting of measurements of drug traces on banknotes seized from drug users. A limited number of studies have analysed drug concentrations on banknotes taken from drug users, but these studies have only analysed a small number of notes (e.g. \cite{ref17}).
    \item Data on drug transfer between surfaces and banknotes or tamper evident bags.
\end{itemize} 
For the last two bullet points it would also be possible to perform experiments to collect data. Given the large number of scenarios considered in the CEG, even if it were possible to carry out the experiments listed above there would still be insufficient data to inform all of the edge probabilities using data alone. Therefore the bulk of the edges in the CEG must be set using expert judgement. This can be combined with sensitivity analysis to see the effect that different choices have on the results. 

Some of the key assumptions regarding the edge probabilities that we have made when developing the CEG are that: 
\begin{itemize}
\item the probability that the suspect had over $\pounds$100,000 cash and then did not bank the notes given that he obtained the notes from a legitimate jewellery business ($H_{d_1}$) is the same as the probability that he had over $\pounds$100,000 cash and did not bank the notes given that he obtained the notes from money laundering but was not aware that this was the case ($H_{d_2}$).
\item the probability that the suspect had over $\pounds$100,000 cash and then did not bank the notes given that he obtained the notes from dealing heroin ($H_{p_1}$) is the same as the probability that the suspect had over $\pounds$100,000 cash and then did not bank the notes given that he obtained the notes from laundering money for drug deals and was aware this was the case ($H_{p_2}$).
\item the probabilities associated with drug use and table contamination are the same under both prosecution hypotheses $H_{p_1}$ and $H_{p_2}$ and under both defence hypotheses $H_{d_1}$ and $H_{d_2}$.
\item the probability density functions associated with the measurements of drug traces on the banknotes and the bags given the various scenarios for drug use and table contamination are the same for $H_{p_1}$, $H_{p_2}$ and $H_{d_2}$ (i.e. the cases where the source of the banknotes was from drug deals). 
\end{itemize}

\subsection{Estimating probabilities of chains of events}

The probability density function directly associated with the forensic evidence (i.e. the measurements of drug traces on the banknotes and the bags) is given by the final edge of the CEG in Figure \ref{ET2}. The likelihood ratio for the full set of prosecution and defence arguments is a function of this probability density function and the probabilities of other edges, which in this example are (almost) entirely defined using expert knowledge. 

The total probability of all chains of events associated with the prosecution arguments in the example case can be calculated using event transition probabilities defined in the chain event graph (Figure \ref{ET2}) as follows:


\begin{equation}
\begin{aligned}
    P(\triangle) = \ & ((1 - \alpha) \ \psi\ + \alpha\ (1 - \beta)\ (1 - \delta)\ \psi) \ \zeta\ (\eta\ (\theta'\ f_7(\mathbf{Y}, \mathbf{Z} \mid s) + \rho'\ f_8(\mathbf{Y}, \mathbf{Z} \mid s) + \tau'\ f_9(\mathbf{Y}, \mathbf{Z} \mid s)) \\
    & + (1 - \eta)\ (\kappa'\ f_{10}(\mathbf{Y}, \mathbf{Z} \mid s) + \upsilon'\ f_{11}(\mathbf{Y}, \mathbf{Z} \mid s) + \phi'\ f_{12}(\mathbf{Y}, \mathbf{Z} \mid s)))
\end{aligned}
\end{equation}




Similarly, the total probability of the chains of events associated with the defence arguments is given by:
\begin{equation}
\begin{aligned}
P(\bigcirc) = \ & \alpha (\beta\ \gamma\ \epsilon\ (\eta \ (\theta\ f_1(\mathbf{Y}, \mathbf{Z} \mid s) + \rho\ f_2(\mathbf{Y}, \mathbf{Z} \mid s) + \tau\ f_3(\mathbf{Y}, \mathbf{Z} \mid s)) + (1 - \eta)\ (\kappa\ f_4(\mathbf{Y}, \mathbf{Z} \mid s) \\ 
& + \phi\ f_5(\mathbf{Y}, \mathbf{Z} \mid s) + \upsilon\ f_6(\mathbf{Y}, \mathbf{Z} \mid s))  + (1 - \beta)\  \delta\ \gamma\ \epsilon\ (\eta(\theta\  f_7(\mathbf{Y}, \mathbf{Z} \mid s) + \rho\ f_8(\mathbf{Y}, \mathbf{Z} \mid s) \\ 
& + \tau\ f_9(\mathbf{Y}, \mathbf{Z} \mid s) + (1 - \eta)\ (\kappa\ f_{10}(\mathbf{Y}, \mathbf{Z} \mid s) + \phi\ f_{11}(\mathbf{Y}, \mathbf{Z} \mid s) + \upsilon\ f_{12}(\mathbf{Y}, \mathbf{Z} \mid s)))
\end{aligned}
\end{equation}

Note that both of these functions are polynomial in the various edge probabilities. This means that the strength of evidence associated with various components of the storyline and the robustness of the different assessments to changing the corresponding edge probabilities can be investigated before these values are known. 

We can also calculate the probability for each individual prosecution and defence storyline in the CEG  $s_{p_i}$ (for $i=\{1,\ldots, 12\}$) and $s_{d_i}$ (for $i=\{1, \ldots, 12\}$) by calculating the product of all event transition probabilities in each individual storyline. For example, $p(s_{d_1})$ is given by:
 \begin{equation}
     p(s_{d_1}) =  \alpha\ \beta\ \gamma\ \epsilon\ \eta\ \theta\ f_1(\mathbf{Y}, \mathbf{Z} \mid s) 
 \end{equation}

\section {Example}

\subsection{Statistical model for the forensic evidence}
\label{section:Bayesian}

Using the chain event graph displayed in Figure \ref{ET2} we formulate a statistical model to evaluate the likelihood ratio associated with the forensic evidence $\mathbf{Y}$ and $\mathbf{Z}$ in the example case under the competing hypotheses proposed by the prosecution and defence. This likelihood ratio is given by:
\begin{equation}
LR = \frac{f(\mathbf{Y}, \mathbf{Z}\mid H_p)}{f(\mathbf{Y}, \mathbf{Z}\mid H_d)}
\label{LR}
\end{equation}
In the case of the prosecution propositions, $H_{p_1}$ is that the suspect was dealing heroin and (${H}_{p_2}$) is that the suspect was knowingly laundering money obtained through the sale of heroin via a jewellery business. The defence proposition $H_d$ is that the suspect either ran a legitimate jewellery business ($H_{d_1}$) or was not aware that his business had been used for the laundering of money obtained through the sale of heroin ($H_{d_2}$).

By conditioning on the relevant parts of the chain event graph (those highlighted in Figure \ref{ET2}), the numerator of the likelihood ratio \eqref{LR} for the prosecution can be written
\begin{align}
f(\mathbf{Y}, \mathbf{Z} \mid H_p)&= \frac{p(H_{p_1})}{p(H_{p_1})+p(H_{p_2})}f(\mathbf{Y}, \mathbf{Z} \mid H_{p_1})+ \frac{p(H_{p_2})}{p(H_{p_1})+p(H_{p_2})}f(\mathbf{Y}, \mathbf{Z} \mid H_{p2}) \nonumber \\
&=w_1\sum_{s_{p_i} \in H_{p_1}} f(\mathbf{Y}, \mathbf{Z} \mid s_{p_i}) \, p'(s_{p_i} \mid H_{p_1})+w_2\sum_{s_{p_i} \in H_{p_2}} f(\mathbf{Y}, \mathbf{Z} \mid s_{p_i}) \, p'(s_{p_i} \mid H_{p_2}) \nonumber \\
&=w_1\sum_{s_{p_i} \in H_{p_1}} f(\mathbf{Y} \mid s_{p_i})f(\mathbf{Z} \mid s_{p_i}) \, p'(s_{p_i} \mid H_{p_1})+w_2\sum_{s_{p_i} \in H_{p_2}} f(\mathbf{Y} \mid s_{p_i})f(\mathbf{Z} \mid s_{p_i}) \, p'(s_{p_i} \mid H_{p_2}) \label{num}
\end{align}

where $p'(s_{p_i} \mid H_{p_j})$ is the probability of storyline $s_{p_i}$ (for $i \in \{1, \ldots, 12\}$ prosecution storylines) conditional on $H_{p_j}$ for $j \in \{1,2\}$ and $w_j$ are weights given by $p(H_{p_j})/(p(H_{p_1})+p(H_{p_2}))$ corresponding to the relative prior probabilities of the two prosecution propositions. Note that we use the notation $p'$ rather than $p$ because these storyline probabilities should be calculated excluding the final edge associated with the forensic evidence. The denominator of the likelihood ratio is obtained in the same way using the propositions $H_{d_j}$ and storylines proposed by the defence $s_{d_i}$. 

Note that \eqref{num} is equivalent to:
\begin{equation*}
f(\mathbf{Y}, \mathbf{Z} \mid H_p)=w_1\sum_{s_{p_i} \in H_{p_1}} p(s_{p_i} \mid H_{p_1})+w_2\sum_{s_{p_i} \in H_{p_2}} p(s_{p_i} \mid H_{p_2}).
\end{equation*}

The second equality sign in \eqref{num} holds because conditional on knowing the contamination status of the table on which both the notes and the tamper evident bags were placed ($T^-$, $TH$, $TA$ in Figure \ref{ET2}) we assume that $\mathbf{Y}$ and $\mathbf{Z}$ are independent. In practice it might be the case that conditional on the table being contaminated with drugs there is still an association between the quantity of contamination of the notes and the number of contaminated bags (this would arise for example if there were a statistical dependence between the quantity of contamination on the table and the number of contaminated tamper evident bags) but we expect this to be a second order effect. A series of experiments examining how surfaces with different quantities of contamination affect the number of contaminated bags and level of contamination on notes could be used to test this assumption. If statistical dependence exists we would expect a positive correlation between the contamination on notes and the number of contaminated bags conditional on the contamination status of the table.

The evidence $\mathbf{Y}$ is a matrix in which each row represents the quantity of drug contamination on each banknote tested (1 to $n$), as measured by the intensity of the $log_{10}$-transformed peak area, and each column, 1 to $m$, represents a different type of drug. We assume that this contamination may have arisen from three possible sources: from drug trafficking; from drug use; or from being counted by the police on a contaminated table. We represent the contamination arising from each of these sources by the following three random variables:
$\mathbf{X}_1$ (trafficking), $\mathbf{X}_2$ (use) and $\mathbf{X}_3$ (contamination from the table).

We model $\mathbf{Y} \mid  \mathbf{s_p}_i$ using a multivariate normal distribution, so that:
\begin{equation}
\mathbf{Y} \mid s_{p_i} \sim MVN(\mathbf{\mu}^{s_{p_i}}, \Sigma^{s_{p_i}}).
\label{distY}
\end{equation}
Here, $\mathbf{Y}$ is conditioned on a given prosecution storyline $s_{p_i}$ (for $i \in \{1, \ldots, 12\}$ storylines) because we expect the parameters of the distribution of $\mathbf{Y}$ to vary depending on the storyline outlined in the chain event graph. We also model $\mathbf{Y} \mid  s_{d_i}$ (for $i \in \{1, \ldots, 12\}$ defence storylines) as a multivariate normal distribution.

Note that as discussed in Section \ref{section:key vectors} we do not have the datasets required to establish a distribution for  $\mathbf{Y} \mid  s_{p_i}$ and  $\mathbf{Y} \mid  s_{d_i}$ empirically. Based on research in \cite{ref12} and \cite{ref13} a Gaussian distribution seems plausible but we use it here as an illustration only. It is possible to modify the analysis described above for non-Gaussian distributions by simply substituting different edge probabilities to the penultimate edges in the CEG consistent with these alternative densities. It is also simple to investigate the robustness of the value of the LR at the observed values of the densities by investigating various other plausible alternative densities. 

To determine estimates of the parameters $\mathbf{\mu}^{s_{p_i}}$ and $\Sigma^{s_{p_i}}$ datasets of drug contamination are needed for each of the possible storylines. Further details are discussed in Section~\ref{section:key vectors}. It may be possible to simplify the problem and reduce the data requirements by establishing a statistical relationship between $\mathbf{Y}, \mathbf{X_1}, \mathbf{X_2}$ and $\mathbf{X_3}$ - e.g. that $\mathbf{Y}\mid \mathbf{s_p}_i=\mathbf{X_1}\mid \mathbf{s_p}_i+\mathbf{X_2}\mid \mathbf{s_p}_i+\mathbf{X_3}\mid \mathbf{s_p}_i$ (conditioning on relevant nodes in the chain event graph) but more experimentation is needed to investigate this relationship. In Section~\ref{section:results} we use illustrative values for the parameters.

As discussed in Section \ref{section:evidence} in this example we will assume that each $z_{ij}$ is a binary variable taking the value of $0$ in the event that the quantity of the $j$-th drug detected on the bag is below the limit of detection of the mass spectrometer and the value of $1$ otherwise. We also assume that conditional on a given storyline, whether or not a bag is contaminated is statistically independent of the contamination status of the other bags tested. Letting $z_j$ be the number of bags contaminated with drug $j$, we can therefore model $z_j$ using a binomial distribution so that: 
\begin{equation}
z_j \mid s_{p_i} = P(z_j;p_j^{s_{p_i}},10) = {10 \choose z_j} (p_j^{s_{p_i}})^{z_j} (1-p_j^{s_{p_i}})^{(10-{z_j})},
\label{distZ}
\end{equation}
where $p_j^{s_{p_i}}$ is the probability of an individual tamper-evident bag testing positive for contamination with the $j$-th drug given $s_{p_i}$. We make a similar assumption for the distribution of $z_j \mid s_{d_i}$. As for $\mathbf{Y}$ it is possible to investigate the sensitivity of the results to both the distributional assumption and the parameter estimate by substituting different possibilities into the final edge probability.

\subsection{Results}
\label{section:results}

As an illustration of how our model can be applied, we generate plausible datasets for the intensity of the $log_{10}$-transformed peak area of a single product ion (m/z 268) representing the quantity of heroin contamination on banknotes and also for the number of tamper-evident bags contaminated with heroin. To simplify the illustration we focus only on the evidence related to heroin contamination, but this could easily be expanded following the description above to a model that includes evidence of contamination with further drugs. We use these generated datasets as hypothetical evidence in a case and use our chain event graph to evaluate the likelihood ratio associated with this evidence. The likelihood ratio is evaluated as described in Section \ref{section:Bayesian}.



To generate vectors of heroin contamination on banknotes ($\mathbf{y}$) for use with the chain event graph, estimates of the parameters $\mu^{s_{p_i}}$,  $\mu^{s_{d_i}}$, $\Sigma^{s_{p_i}}$ and $\Sigma^{s_{d_i}}$ as in equation \eqref{distY} are needed for each storyline. To generate the number of contaminated tamper evident bags ($z$), estimates of $p^{s_{p_i}}$ and $p^{s_{d_i}}$ as in equation \eqref{distZ} are needed for each storyline. Note that as we are only considering heroin, the subscript $j$ representing the particular drug has been dropped. The values we used for these parameters are shown in the fifth column of Table \ref{table:Table3}, where the first number is $\mu$, the second $\Sigma$ and the third $p$.

As discussed above, we have assumed that the heroin contamination on the banknotes (above that found as background) comes from three possible sources: trafficking, drug use and table contamination. As a result there are 12 possible sets of parameter values for $\mu$ and $\Sigma$ corresponding to the 12 possible combinations of these three sources (shown in Table \ref{table:Table3}). As there are more than 12 possible storylines in the chain event graph this means that some of the storylines are associated with the same distribution of heroin contamination on the banknotes (i.e. the same final edge probability). This is shown by the 12 probability distributions $f_1$ to $f_{12}$ marked on the final edges in the chain event graph. Note that while $f_1$ to $f_6$ are associated with storylines proposed by the defence, $f_7$ to $f_{12}$ are associated with both defence and prosecution storylines. This is because the prosecution proposition $H_{p_2}$ and the defence proposition $H_{d_2}$ differ only in whether the suspect did or did not know that the notes had been involved in drug trafficking. The parameter $p$ (the probability of a tamper evident bag testing positive for drug contamination under a given storyline) depends only on the table contamination. 

The parameters in each row of Table \ref{table:Table3} were used to generate sets of data $\mathbf{y}$ and $z$ from the normal and binomial distributions given in equations \eqref{distY} and \eqref{distZ} respectively. Then, for each of these datasets, the likelihood ratio was computed. That is, we determined what the likelihood ratio would be if the parameters in each row of the table represented the true distribution of contamination on the banknotes and the tamper evident bags. The likelihood ratio is given by the ratio of $f(\mathbf{y},z|H_p)$ and $f(\mathbf{y},z|H_d)$, computed using equation \eqref{num}. 

To compute the likelihood ratio, the probabilities $p'(s_{p_i})$ and $p'(s_{d_i})$ conditional on the various propositions are needed. These probabilities are calculated as the product of edges in the chain event graph, where the illustrative edge probabilities are given in Table \ref{table:Table4}. For example, 
\begin{equation}
    p'(s_{d_1} \mid H_{d_1}) =  \gamma \epsilon \eta \theta .
\end{equation}

For each row in the table, contamination levels were simulated for 50 banknotes and for 10 tamper-evident bags. The simulation was repeated 100 times for each row. The mean log-likelihood ratio over these 100 simulations is displayed in the far right column of Table \ref{table:Table3}. As expected, when datasets are not associated with heroin trafficking the log-likelihood ratio is large and negative, decreasing slightly when there is heavy drug use. When there is heroin trafficking, the log-likelihood ratios are positive. In this latter case, the possibility that the suspect did not know about the heroin trafficking (a defence proposition) has been included in the calculations, but as the edge probabilities $1-\beta$ and $\delta$ are small (representing the storylines where the jewellery business is being used to launder money from a drug trafficking operation but the suspect is unaware) the likelihood ratios are still high. This is despite the fact that the evidence itself (the drug contamination on the notes and on the tamper evident bags) is the same whether the suspect did or did not know about the drug trafficking. In this way the incorporation of edge probabilities representing activities in the case has enabled us to make a probabilistic assessment that is supportive of the prosecution. If modelling only the forensic evidence (and not the activities) the likelihood ratio of $H_{p_2}$ against $H_{d_2}$ would be one because the evidence could arise both under a defence and a prosecution proposition. 
 
\begin{table}
    \centering
    \begin{tabular}{|l|lll|cc|}
    \hline
   Prob. & Heroin& Drug use& Table& Parameter& log-LR \\
    Dist&  trafficking?&&contamination&values&\\
     \hline
      $f_1$&      N   & Heroin only &No heroin&4.35, 0.38, 0.05&-41.2\\
  $f_2$&     N   & Heroin only &Heroin only&4.4, 0.38,0.45&-40.0\\
  $f_3$&     N   & Heroin only &Heroin and other drugs&4.45, 0.38, 0.55&-34.9\\
  $f_4$&     N   & Heroin and other drugs &No heroin&4.5, 0.4, 0.05&-22.6\\
   $f_5$&    N   &  Heroin and other drugs&Heroin only&4.55,0.4, 0.45&-21.3\\
   $f_6$&    N   &  Heroin and other drugs&Heroin and other drugs&4.6, 0.4, 0.55&-17.0\\
    $f_7$&  Y   & Heroin only &No heroin&5, 0.38, 0.05&8.3\\
     $f_8$&   Y  & Heroin only &Heroin only&5.05, 0.38, 0.45&8.8\\
    $f_9$&     Y & Heroin only &Heroin and other drugs&5.1, 0.38, 0.55&8.8\\
  $f_{10}$&   Y     & Heroin and other drugs&No heroin&4.95, 0.45,0.05&8.4\\
  $f_{11}$&    Y    &  Heroin and other drugs&Heroin only&5, 0.45, 0.45&9.0\\
  $f_{12}$&     Y   &  Heroin and other drugs&Heroin and other drugs&5.05, 0.45, 0.55&8.9\\
      \hline
    \end{tabular}
    \caption {Parameter values assumed for final evidence edge in CEG (corresponding to labels in \ref{ET2}) and log-likelihood ratios for evidence data simulated from these parameter values.}
    \label{table:Table3}
\end{table}

\begin{table}
    \centering
    \begin{tabular}{|lc|lc|lc|lc|}
    \hline
   $\alpha$&0.9&$\beta$&0.9&$\delta$&0.1&$\gamma$&0.05\\
   $\epsilon$&0.01&$\eta$&0.5&$\eta'$&0.4&$\theta$&0.45\\
   $\rho$&0.45&$\kappa$&0.33&$\upsilon$&0.33&$\psi$&0.8\\
   $\zeta$&0.9&$\theta'$&0.3&$\rho'$&0.6&$\kappa'$&0.2\\
   $\upsilon'$&0.4&&&&&&\\
     \hline
\end{tabular}
\caption{Assumed edge probabilities for the example}
    \label{table:Table4}

\end{table}

In the Appendix (Figure \ref{fig:sens}) we present a sensitivity analysis for data generated from the parameters in rows 6 (left hand side) and 7 (right hand side) of Table \ref{table:Table3}. A sensitivity analysis like this could be used to investigate whether any of the subjectively set edge probabilities are particularly dominant in determining the likelihood ratio. We can see from the plots that in the case where the data are generated from a distribution that assumes no drug trafficking (left hand side), the log-likelihood ratios are always negative and large except in the case where both $\beta$ and $\alpha$ are small. This corresponds to a high probability that the suspect was dealing heroin and that in the case that they were running a jewellery business, it was involved in money laundering so this result is expected. When the data are generated from a distribution that assumes there was drug trafficking there are more situations in which the log-likelihood ratios are not conclusive. For example, when there is a high probability that money laundering was occurring but that the suspect did not know about it (small $\beta$, large $\delta$). 

\section{Discussion/Conclusion}

The ability to assess activity-level propositions is important in cases concerning trace evidence such as the drug trafficking example studied in this paper (\cite{ref3}). In the example studied the main forensic evidence was the quantity of drug contamination found on banknotes. The source of this contamination was unknown and could have come from various sources (e.g. drug trafficking activities, drug use by suspect or those in contact with the suspect, contamination from police activities). Understanding the mechanisms by which the drug contamination came to be on the notes was a crucial aspect considered in the trial. 

One problem with determining the value of evidence where this evidence relates to drug contamination on surfaces is that it is necessary to understand what actions by the suspect resulted in this observed contamination. To do this often requires consideration of how multiple different pieces of evidence might support different possible explanations for the drug contamination. Such evidence might include testimony or expert knowledge about the behaviour of drug traffickers, as well as contamination measurements on different surfaces.

Statistical tools which evaluate expert knowledge along with other evidence from a case (such as quantity of contamination on banknotes) are already available. Previous studies have used Bayesian networks to evaluate multiple different pieces of evidence and to address activity-level propositions in forensic science (\cite{ref18}, \cite{ref4}, \cite{ref22}). However, Bayesian networks have several disadvantages when used for forensic cases, mainly that they are not intuitive in their representation of temporal information. To compare probabilities of different event sequences, other graphical methods may be more useful. Chain event graphs (\cite{ref10}) are graphical probability models for comparing event sequences and have been utilised in various fields for this purpose, including in healthcare (\cite{ref19}, \cite{ref20}) and epidemiology (\cite{ref21}). 

In this paper we developed a statistical framework using chain event graphs to evaluate evidence for activity-level propositions in a drug trafficking case in which possible activities of the suspect are compared using multiple pieces of available evidence as well as expert knowledge. We also demonstrated how this chain event graph might be used to estimate the likelihood ratio for competing activity level propositions associated with the forensic evidence in the case. We introduced two novel modifications to the standard chain event graph to make it more suitable for use with forensic evidence. Firstly we depicted the chain event graph after conditioning on evidence that had been accepted by all parties. This greatly reduces the complexity of the chain event graph and allows us to include continuous forensic evidence in the calculations. Secondly we used two sink nodes rather than one to collect the prosecution root-to-leaf paths together and the defence root-to-leaf paths together. This makes the graph more intuitive to those used to the adversarial system.

Each event transition in the chain event graph shown in Figure \ref{ET2} has an associated probability (i.e. probability of moving from one event to the next in a storyline). In examples in which few relevant data are available, estimating these event probabilities will rely heavily upon expert elicitation. The use of expert judgement can be a valuable addition to other forms of evidence in decision making. However, how much confidence can be placed on these judgements is debatable and is likely to depend on the choice of expert(s) and the availability of data supporting their opinions. In our example case, we expect that most event transition probabilities will be based solely on expert judgement. We demonstrated how a sensitivity analysis can be used to assess which event transition probabilities in the CEG are more influential in affecting the likelihood ratio. A chain event graph enables those involved in a case to set out the arguments and evidence in a logical and clear manner and to assess which assumptions might have the largest effect on the conclusions. If a graphical method were not used these arguments and expert judgements would still need to be made to assess the evidence presented but they may not be as explicit and hence as open to criticism and testing. 

Our analysis may be improved by identifying or collecting relevant data to include in the CEG or inform the event transition probabilities. There are also other possible arguments and types of evidence that we did not consider in our chain event graph. For example, we did not consider the pattern of banknotes contaminated with drugs within a bundle of notes. Although we did not make use of this evidence in our study, this information could potentially be used to differentiate between contamination from handling by the police vs contamination by the suspect (through his own drug use or from drug trafficking). We would expect variation in contamination among notes obtained through drug trafficking to be greater than variation among notes contaminated from the suspect's own drug use, as notes from different customers are likely to have different levels of contamination depending on the customer's own level of drug use. Similarly, the pattern of contamination on notes within bundles could be used to differentiate between police contamination and contamination from other sources as pattern of contamination on notes within a bundle would be expected to reflect the counting pattern undertaken by police if contamination were from this source.

This paper demonstrates how a chain event graph can be used to compare competing versions of events proposed by the prosecution and defence in a criminal case. A chain event graph can display multiple different arguments along with multiple different pieces of evidence and be used to perform likelihood ratio calculations for forensic evidence. This is crucial in cases which involve activity-level propositions. Our example based on a real-world drug trafficking case shows how our methodology can be used to combine and assess different types of evidence allowing activity-level propositions to be compared.

\section*{Acknowledgements}

Author Amy Wilson thanks the late Dr Richard Sleeman for support throughout her career so far in forensic science and for many interesting exchanges on the topic of drugs on banknote evidence. 

All of the authors gratefully acknowledge funding from the Alan Turing Institute under wave one of the UKRI Strategic Priorities Fund, EPSRC Grant EP/W006022/1, particularly the “Criminal Justice System” theme within that grant. JQS and ALW also acknowledge Turing Fellowships from the Turing Institute.

\newpage
\bibliographystyle{apa}
\bibliography{references}

\section*{Appendix}

\begin{figure}[H]
\centering
\includegraphics[width=0.7\linewidth]{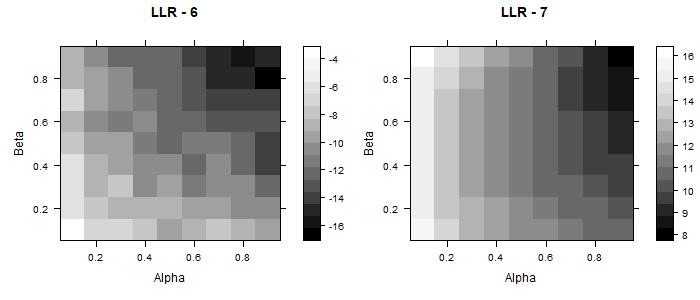}\\
\includegraphics[width=0.7\linewidth]{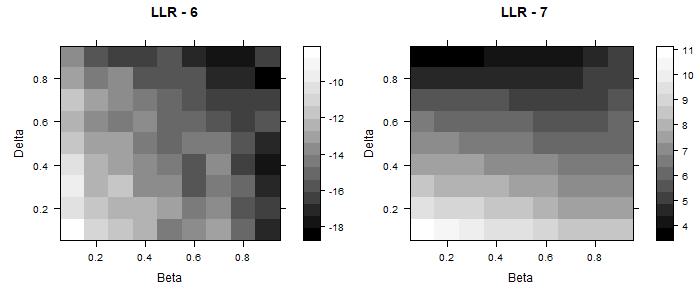}\\
\includegraphics[width=0.7\linewidth]{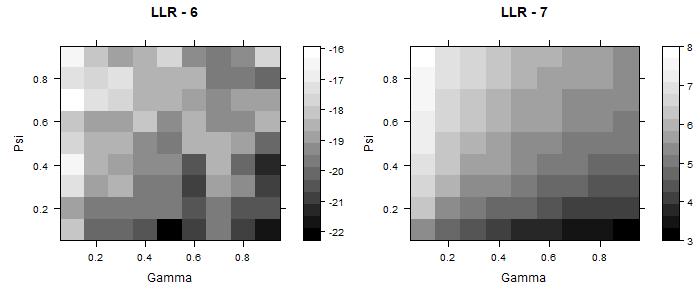}\\
\includegraphics[width=0.7\linewidth]{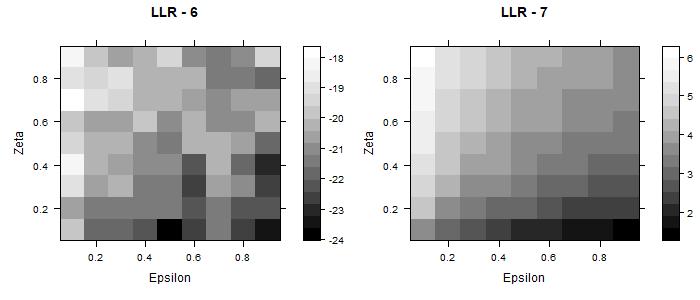}\\
\end{figure}

\begin{figure}[H]
\centering
\includegraphics[width=0.7\linewidth]{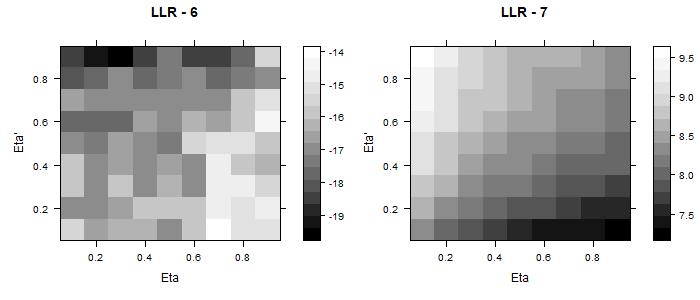}\\
\includegraphics[width=0.7\linewidth]{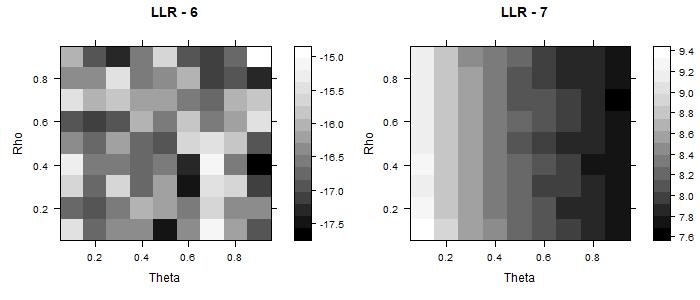}\\
\includegraphics[width=0.7\linewidth]{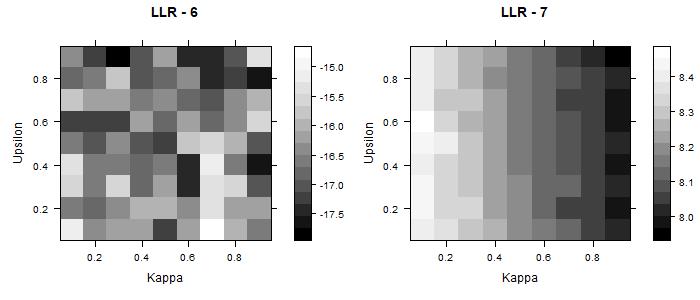}\\
\includegraphics[width=0.7\linewidth]{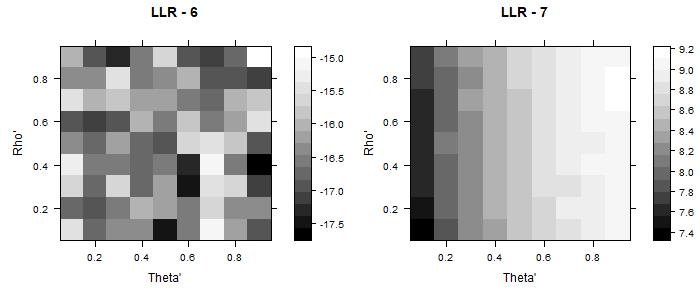}\\
\end{figure}

\begin{figure}[H]
\centering
\includegraphics[width=0.7\linewidth]{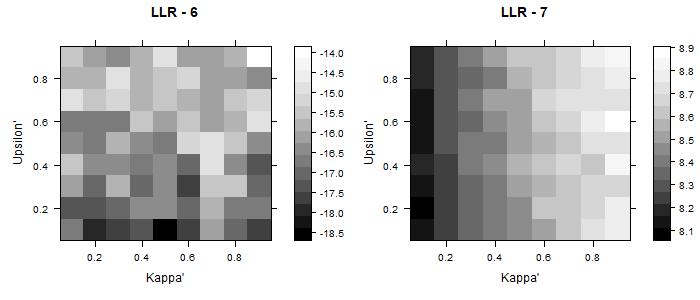}\\
\caption{Sensitivity analysis - the effect of varying edge probabilities on the log-likelihood ratio for two different simulated sets of banknote evidence. The left hand side shows results corresponding to row 6 in Table \ref{table:Table3} and the right hand side shows results corresponding to row 7 in Table \ref{table:Table3}}
\label{fig:sens}
\end{figure}

\end{document}